\documentclass[useAMS,usenatbib]{mn2e}
\usepackage{graphicx}
\usepackage{times}
\usepackage{epsfig}
\usepackage{longtable}
\usepackage{lscape}
\usepackage{amssymb}
\usepackage{latexsym}
\usepackage{natbib}
\def\kms{$\rm km\;s^{-1}$}

\def\hb{H$\beta$}

\def\mguno{Mg$_1$}
\def\mgdue{Mg$_2$}
\def\mgb{Mg{\it b}}
\def\mgi{Mg{\small I}}
\def\fei{Fe{\small 5270}}
\def\feii{Fe{\small 5335}}

\title[Stellar populations and formation history of the BCG NGC 4889]
{Distinct core and halo stellar populations and the formation history of
  the bright Coma cluster early-type galaxy NGC 4889}

\author[L. Coccato et al.]{Lodovico Coccato$^{1}$\thanks{E-mail: lcoccato@mpe.mpg.de, gerhard@mpe.mpg.de},
  Ortwin Gerhard$^{1}$, Magda Arnaboldi$^{2,3}$  \\
  $^{1}$Max-Planck-Institut f\"ur extraterrestrische Physik, D-85748 
          Garching b. M\"unchen, Germany\\
  $^{2}$European Southern Observatory, Karl-Schwarzschild-Stra$\beta$e 2, D-85748 
          Garching bei M\"unchen, Germany\\
  $^{3}$INAF, Osservatorio Astronomico di Pino Torinese, I-10025 Pino
       Torinese, Italy
}
\begin{document}

\date{June 4, 2010}  

\pagerange{L\pageref{firstpage}--L\pageref{lastpage}} \pubyear{2010}

\maketitle

\label{firstpage}

\begin{abstract}
  We study the stellar population far into the halo of one of the
    two brightest galaxies in the Coma cluster, NGC 4889, based on
  deep medium resolution spectroscopy with FOCAS at the Subaru 8.2m
  telescope.  We fit single stellar population models to the measured
  line-strength (Lick) indices (\hb, \mgb, [MgFe]' and $<$Fe$>$).
  Combining with literature data, we construct radial profiles of
  metallicity, [$\alpha$/Fe] element abundance ratio and age for
  NGC~4889, from the center out to $\sim$ 60 kpc ($\sim4 R_e$).
  We find evidence for different chemical and star formation histories
  for stars inside and outside $1.2 R_e = 18$ kpc radius.  The inner
  regions are characterized by a steep [Z/H] gradient and high
  [$\alpha$/Fe] at $\sim$2.5 times solar value. In the halo, between
  $18$ and $60$ kpc, the [Z/H] is near-solar with a shallow
  gradient, while [$\alpha$/Fe] shows a strong negative gradient,
  reaching solar values at $60$ kpc.
  We interpret these data in terms of different formation histories
  for both components. The data for the inner galaxy are consistent
  with a rapid, quasi-monolithic, dissipative merger origin at early
  redshifts, followed by one or at most a few dry mergers. Those for
  the halo argue for later accretion of stars from old systems with
  more extended star formation histories.  The half-light radius of
  the inner component alone is estimated as $\sim 6$ kpc, suggesting a
  significantly smaller size of this galaxy in the past. This may be
  the local stellar population signature of the size evolution found
  for early-type galaxies from high-redshift observations.
  \end{abstract}


\section{Introduction}
\label{sec:introduction}

Brightest cluster galaxies (BCGs) are the largest and most luminous
galaxies located in galaxy clusters.  The merger, star formation and
chemical enrichment history of these galaxies are imprinted in their
kinematics and chemical abundances. Spatially extended measurements
for such quantities can therefore constrain their evolution and
formation, believed to be closely related to the formation of the
cluster (e.g., \citealt{Dubinski98}) and the presence of the diffuse
intra-cluster light (e.g., \citealt{Napolitano+03,Murante+07}).

In the cold dark matter-based scenario, early-type galaxies (ETGs)
form via merging of subclumps with various masses. According to
\citet{Kobayashi04}, their merging histories can vary between
spherical infall of gas-rich subunits at high redshift, and a sequence
of merging events at different epochs and masses. Galaxies of the
former assembly history are created via a process that is similar to
the classical monolithic collapse. This dissipative process gives rise
to high metallicity in the galaxy center, and to significant
logarithmic metallicity ([Z/H]) gradients that can be steeper than
$-0.5$ and correlate strongly with galaxy mass (e.g.,
\citealt{Chiosi+02, Kobayashi04, Pipino+08}).  Gas-rich binary
mergers, on the other hand, produce only shallow gradients ($\sim
-0.1$) with weak mass dependence \citep{Bekki+99}.  Merging events
that take place after most of the stars are in place reduce, but do
not erase completely the pre-existing gradients, because of the way
in which stars of different metallicities are redistributed in both
energy and radius \citep[][]{White80,diMatteo+09}.  In case of a later
gas rich merger with a central starburst, a metallicity gradient can
form again, but is confined to the central regions after a few Gyr
\citep{hopkins+09}.

Stellar population studies in ETGs based on spectral indices show
metallicities higher than solar in their nuclei, negative logarithmic
metallicity ([Z/H]) gradients ranging from $-0.16$ to $-0.30$, and
nearly constant [$\alpha$/Fe] element abundances with radius
(e.g., \citealt{Kobayashi+99, Sanchez-Blazquez+06b, Reda+07,
  Annibali+07}). For BCGs, several studies provide information on the
chemical abundances reaching up to $\sim 20$ kpc radius
\citep{Carter+99, Brough+07, Loubser+09}. No significant gradients of
[$\alpha$/Fe] are observed, as in the case of normal
early-type galaxies, while the range of [Z/H] gradients is
somewhat wider, with values from $-0.20$ down to $-0.58$. Stellar
ages are determined to be mostly old, $>8$~Gyr.

In this paper we analyse Lick absorption line indices from deep
  spectra for NGC~4889, one of the two central BCGs in the Coma
  cluster \citep[see, e.g.][]{Gerhard+07}, reaching unprecedented
radii of $\sim 60$ kpc ($\sim$ 4 $R_e$) in its outer halo (Section
2). We then derive age, [Z/H] and [$\alpha$/Fe] radial distributions
using single stellar population models (Section 3).  Finally, we
discuss the implication of these stellar populations properties for
the formation history of the core and halo of this galaxy
(Section~\ref{sec:discussion}). In what follows we adopt a distance of
$D=92.7$ Mpc (NASA/IPAC Extragalactic Database) and effective radius
$R_e = 33\farcs88$ = 15.23 kpc \citep{Jorgensen+95} for NGC~4889.

\section{Line-strength index profiles for NGC 4889}
\label{sec:observations}

\begin{figure}
\psfig{file=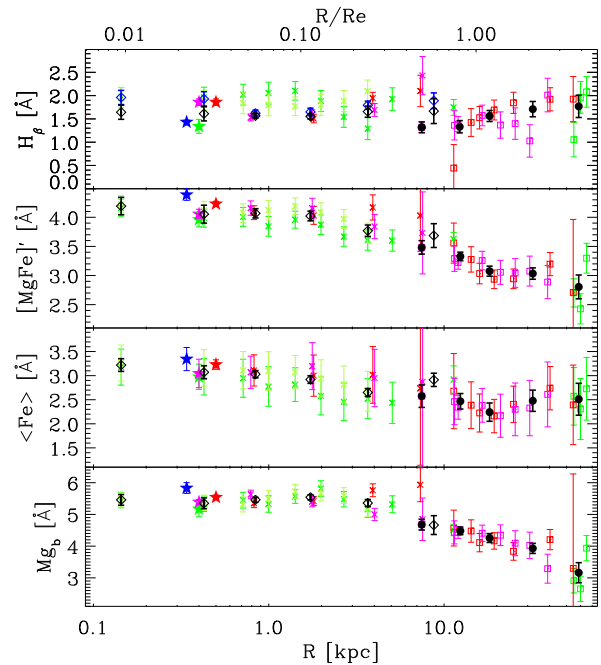,width=8.3cm,clip=}
\caption{Lick absorption line indices for NGC~4889 as projected onto
  the galaxy major axis. {\it Diamonds:} data from \citet{Mehlert+00,
    Corsini+08}, averaged in radial bins. Here the second panel gives
  unprimed [MgFe], and H$\beta$ values are shown without (blue) and
  with offset correction (black); see text. {\it Filled circles:}
  outer halo data from Paper I, also averaged in radial bins. Second
  panel gives primed [MgFe]'; see Table~\ref{tab:models}. {\it Squares
    and crosses:} measurements before averaging (dark and light green
  are from both sides of major axis spectrum, red and pink from both
  sides of minor axis spectrum).  {\it Stars:} Central aperture
  measurements, from \citet[][red]{Jorgensen99},
  \citet[][magenta]{Moore+02}, \citet[][green]{Sanchez-Blazquez+06a},
  \citet[][blue]{Loubser+09}.  }
\label{fig:indices}
\end{figure}

Combining new data for the outer halo of NGC 4889 with literature data
we first construct radial Lick index profiles extending from the
center to $\sim$ 60 kpc radius ($\sim4 R_e$); see
Figure~\ref{fig:indices}.  

The outer halo measurements are taken from \citet[paper I]{Coccato+10}. 
These data cover the radius range $\sim 7$ kpc to $60$ kpc and
  are based on 8 hr (run 1, PA=$81^{\circ}7$, $\sigma_{\rm inst}=76$
  \kms) and 5.5 hr (run 2, PA=$-8^{\circ}3$, $\sigma_{\rm inst}=96$
  \kms) exposures, long-slit spectra obtained with FOCAS@Subaru on
  Mauna Kea. Typical signal to noise ratios range from $\sim$60 to
  $\sim$10 for the innermost and outermost spectra, respectively.

The data analysis is described in Paper I and includes wavelength-
and flux-calibration, convolution to the spectral resolution of the
Lick system, correction for line-of-sight velocity broadening and
correction for the offset to the Lick system.
The sky background is evaluated from both offset sky exposures
  and dark regions in the scientific frames, using the former to
  identify regions of the slit in the latter that are free from
  stellar light continuum. Then we used the sky spectra from these
  regions ($\sim$5\farcm6 from the galaxy center in run $1$, and
  $\sim$2\farcm9 in run $2$); they have the same continuum as in the
  offset sky, and ensure the best removal of the variable sky emission
  lines.  After sky subtraction, residual unresolved emission lines
  are still present in the \mgi\ bandpass (run 2) and H$\beta$ blue
  pseudo continuum (run 1). They are possibly caused by internal
  reflection in the spectrograph and are removed via a Gaussian fit
  (Paper I).
  Scattered light is measured in regions of the CCDs not illuminated
  by the slit. It amounts to $\sim$ 4.5--5.0 ADUs ($\sim$20\% of sky),
  also in the offset sky images, and is nearly uniform across the
  detectors in both runs; it is subtracted off the science frames
  together with the sky background.

On the final spectra, \hb, \mguno, \mgdue, \mgb, \fei\ and \feii\
line-strength indices are measured \citep[see][]{Worthey+94}, from
which we determine $<$Fe$>$ = $({\rm \fei} + {\rm\feii})/2$ and
[MgFe]' = $\sqrt{{\rm \mgb}\left(0.72\cdot{\rm \fei}+0.28\cdot{\rm
      \feii}\right)}$. Their errors are determined by Monte Carlo
simulations which accounted also for the errors in the radial
velocities.
Special care is taken to quantify possible systematic
errors caused by sky subtraction.
Residual contributions of up to $\pm$2\% of the subtracted sky
  spectrum are added to the galaxy spectrum. Larger residuals are
  excluded because they would produce detectable line features in the
  spectra (see Paper I). New index values are then determined for
  these new spectra. The systematic deviations of the new values are
  found to be no larger than the error bars, except for \mgdue, which
  we therefore do not use to derive stellar population parameters.

For the inner regions of NGC 4889, we use spatially resolved Lick
index measurements along the major axis from \citet{Mehlert+00}, and
along the minor axes from \citet{Corsini+08}.  These authors provide
[MgFe]=$\sqrt{{\rm \mgb <Fe>}}$ instead of [MgFe]'.  We also include
central aperture measurements from \citet{Jorgensen99,Moore+02,
  Sanchez-Blazquez+06a,Loubser+09}.  For each index, all halo and
inner data are combined in a single radial profile (see
Fig.~\ref{fig:indices}), by projecting the minor axis data points onto
the major axis using constant ellipticity $\epsilon=0.34$
\citep{Bender+89}.

Average profiles are constructed by grouping and averaging the data
points in radial bins.  For the halo data, we use 5 radial bins
centered at 7, 12, 18, 32 and 59 kpc from the galaxy center. The inner
literature data are similarly grouped in 6 bins, with average radii
0.14, 0.43, 0.84, 1.7, 3.7, and 8.7 kpc.  Errors for the average
values are determined from the errors in the individual points and
from the scatter in the bins. This scatter derives partially from
observational errors, but also from systematic differences between
major and minor axis and between both sides of the same long-slit
spectrum. To take these systematic effects into account, we compute
the $\chi^2$ of the weighted mean with respect to the error bars of
the $m_{\rm bin}$ individual points in each bin. In case it is larger
than $m_{\rm bin}$, the error of the weighted mean is increased
accordingly.

Within the final errors, the inner and outer datasets agree in the
region of overlap ($\sim$7-15 kpc), except in \hb\ where the more
accurate newer data give lower values.  \citet{Trager+08} show that
the H$\beta$ linestrengths measured by \citet{Mehlert+00} for several
Coma galaxies including the second cD galaxy NGC 4874 are stronger by
an average 0.38\AA\ than all other literature values, possibly due to a
too short slit leading to incorrect sky subtraction.  For NGC 4889 we
also find that the H$\beta$ aperture value determined by
\citet{Mehlert+03} is larger by 0.32\AA\ than the four aperture
measurements shown in Fig.~\ref{fig:indices}, while the metal indices
agree within the errors. We therefore construct a second,
offset-corrected set of inner average points for H$\beta$ only, using
the data from \citet[][corrected downward by 0.32\AA\ ]{Mehlert+00} and
\citet[][uncorrected]{Corsini+08}.  These averaged radial profiles are
shown in Fig.~\ref{fig:indices} along with the individual
measurements.

\section{Single stellar population (SSP)
 parameters: [Z/H], [$\alpha$/Fe] and age
}
\label{sec:models}

In all bins of the radial average profiles in Fig.~\ref{fig:indices},
we determine luminosity-weighted metallicity [Z/H], [$\alpha$/Fe]
abundance ratio, and population age from fitting single stellar
population models of \citet{Thomas+03} to the line-strength indices
\hb, \mgb, $<$Fe$>$ and [MgFe]' or [MgFe].
Their grid of models covers age values from 0.1 to 15
Gyr, [Z/H] from $-2.25$ to $0.67$ and [$\alpha$/Fe] from $-0.3$ to
$0.5$.  We interpolated this model grid so that we have steps of
$\sim$ 0.4 Gyr in age, $\sim 0.018$ dex in [Z/H] and $\sim 0.02$ in
[$\alpha$/Fe]. For one of the central data points and a few 
error bars we had to extrapolate to metallicities up to 0.9.

For each radial bin, we determine the best fit model from the measured
\hb, \mgb\, [MgFe]' and $<$Fe$>$ values by minimizing $\chi^2$. Errors
for [Z/H], [$\alpha$/Fe] and ages are computed by means of Monte Carlo
simulations.  The results are shown in Figure \ref{fig:models} and are
listed in Table~\ref{tab:models}.  
Logarithmic gradients of the SSP parameters are
determined from standard linear regression fits in log-log-diagrams.
For given parameter $P$ (denoting either [Z/H], [$\alpha$/Fe],
or $\log_{10}[{\rm age/Gyr}]$), its logarithmic gradient $\Delta P$ is
defined through $ P(R) - P(R_e) = \Delta P \cdot \log_{10} \left(
  R/R_e \right)$.  The derived stellar population gradients are shown
in Fig.~\ref{fig:models}.

\subsection{Results}
\label{sec:results}

\begin{figure}
 \psfig{file=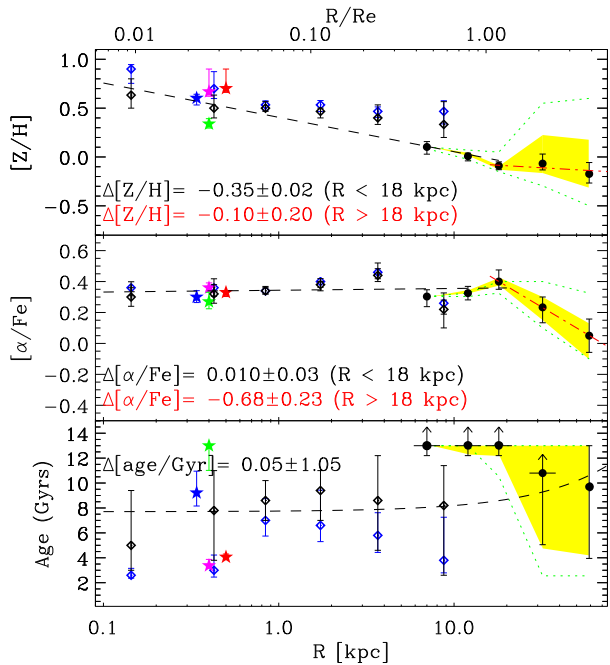,width=8.3cm,clip=}
\caption{Single stellar population parameters and their gradients in
   NGC 4889. {\it Diamonds} show values derived from the data of
   \citet{Mehlert+00,Corsini+08} averaged in radial bins, without (blue)
   and with offset correction (black); see Fig.~\ref{fig:indices}.  
   {\it Filled circles} show halo
   values derived from the data of Paper I also averaged in radial
   bins; see Table~\ref{tab:models}. {\it Stars} show values derived
   from the aperture measurements given in Fig.~\ref{fig:indices}.
   The {\it dashed} lines show the logarithmic gradients in [Z/H], [$\alpha$/Fe], and age
   fitted to these data (in two radial ranges for the first two), with numerical
   values given on the figure. 
   The {\it shaded} yellow regions indicate the range of SSP
     parameters that would result if systematic residual effects in
     the sky subtraction at the level of $\leq \pm$2\% were
     present. The {\it green} lines correspond to (unrealistic)
     $\pm$8\% residuals (see Section \ref{sec:results} and paper I). 
 }
\label{fig:models}
\end{figure}

Over the entire radial range from the centre to $60$ kpc, the
metallicity and [$\alpha$/Fe] profiles in Fig.~\ref{fig:models} can no
longer be described as single power-laws. Rather, they show a
``break'' at $\sim 18$ kpc ($\sim$1.2 $R_e$): in the inner regions,
the logarithmic [Z/H] gradient is negative and steep:
$\Delta[Z/H]\!=\!-0.35\pm 0.02$ ($-0.49 \pm 0.03$) using the corrected
(uncorrected) inner H$\beta$ data, with metallicity that decreases
from $\sim$5 times solar in the centre to $\sim$0.8 solar at $18$
kpc. The [$\alpha$/Fe] abundance ratio in the inner 1.2$R_e$ is nearly
constant at a value of $0.37$, i.e., $\sim$2.5 times solar.  By
contrast, in the halo outside $\sim$ 1.2 $R_e$, the logarithmic [Z/H]
gradient flattens, $\Delta[Z/H] = -0.1\pm0.2$, and the [$\alpha$/Fe]
profile shows a steep negative gradient, $\Delta$[$\alpha$/Fe]$ =
-0.68\pm0.23$, reaching solar values at $\sim$60 kpc. No peculiar
  features corresponding to this ''break'' are seen in the surface
  brightness profile \citep{Thomas+07}.

The robustness of the ``break'' in the abundance profiles to
  possible systematic errors caused by sky subtraction has been
  quantified as follows.  New SSP models are computed using Lick
  indices derived from spectra including up to $\pm$2\% residual sky
  contribution (Section \ref{sec:observations}). The corresponding
  range of SSP parameters is shown by the yellow regions in
  Fig. \ref{fig:models} and approximately corresponds to the detection
  limit for the ``break'' in the [Z/H] radial profile. Residuals of
  (unrealistic) $\pm$8\% of the subtracted sky spectrum would be
  required to remove also the ``break'' in [$\alpha$/Fe] (green lines
  in Fig. \ref{fig:models}).

The inferred SSP ages show an almost bimodal pattern.  The halo of
NGC~4889 at radii $>\!18$ kpc is uniformly old, with SSP ages ranging
between $10$ and $13$~Gyr.  Younger ages are determined for the inner
regions of NGC~4889: using the original H$\beta$ values results in
average SSP ages between $2$ and $8$~Gyr, with the lowest values
inferred in the central kpc. With the offset correction to the average
literature H$\beta$ scale, average ages in the inner regions increase
to $8$~Gyr, but some of the aperture data still imply quite young ages
($\sim3$~Gyr; Fig.~\ref{fig:models}).

Young inferred SSP ages for BCG centers are not uncommon
(\citealt{Fisher+95}, \citealt{Pipino+09}).  However, such young
  SSP ages, if real for NGC 4889 and not caused by the heterogeneous
  nature of the inner measurements, do not necessarily indicate that
  the main bulk of stars are young but can also be due to a
  contamination of a small percentage of young stars. In fact, a
component with age 1 Gyr and 10\% mass fraction can bring the inferred
SSP age of an otherwise old population down to $\sim$3 Gyr
(\citealt[][Table 8]{Trager+00b}), similar to the lowest central
values in NGC 4889. Conversely, the inferred SSP [Z/H] and
[$\alpha$/Fe] of a composite stellar population follow approximately
its V-band luminosity weighted values \citep{Serra+Trager07}, i.e.,
are only moderately sensitive to the [Z/H] of a minor young component
(correspondingly less than the difference between the black and blue
points in the top panel of Fig.~\ref{fig:models}). Thus the steep
[Z/H] gradient and high [$\alpha$/Fe] characterize the dominant
stellar population in the inner half of NGC 4889.

\section{Discussion}
\label{sec:discussion}

From measurements at large radii combined with previous literature
data, we find evidence for distinct stellar populations in the inner
half and outer halo of the nearby BCG NGC 4889. These point to a rapid
formation process for the core and a more extended accretion phase
building the halo, as we now discuss.

Within $\sim1.2 R_e=18$ kpc, all [$\alpha$/Fe] measurements give high
values, implying a flat profile at $\sim 2.5$ times solar abundance
ratio.  This implies rapid conversion of the gas into stars. Using a
formula derived by \citet{Thomas+05} from simple chemical evolution
models would predict that most of the stars within 18 kpc radius were
formed in less than 0.1 Gyr. This is about one dynamical time at 18
kpc. However, given the current uncertainties about SNIa progenitors
and delay time distributions \citep[e.g.][]{Pritchet+08}, this
timescale of 0.1 Gyr is not very secure.
At the same time, the combined measurements support a steep inner
[Z/H] gradient (at least $\Delta[Z/H]\sim -0.3$ dex), while the
published SSP age values inside 18 kpc show considerable scatter,
including both old (10-13 Gyr) and intermediate age (3-6 Gyr) values.
These data are consistent with a superposition of a dominant old
population with a minor, younger component (see previous section).

The rapid formation and steep [Z/H] gradient of the stars in the inner
half of NGC 4889 are reminiscent of predictions of a quasi-monolithic
dissipative collapse model, in which stars form in a rapid burst
progressing from the outside in, while the gas collapsing to the
center is continuously enriched
\citep{Carlberg84,Arimoto+87,Thomas+99}. In this process, logarithmic
gradients of $\sim-0.5$ can be reached.
The chemical properties of the inner NGC 4889 population are
consistent with a scenario in which a dominant, old population formed
rapidly at high redshift in quasi-monolithic, dissipative merger
collapse, as described by \citet{Kobayashi04} in the context of
hierarchical models. Subsequently, several of such units could have
been involved in one or at most a few dry mergers, such that the
original steep [Z/H] gradients are only partially erased - a
major dry merger between two ellipticals with identical gradients
already reduces the gradient by a factor 0.6 \citep{diMatteo+09} - and
such that at most a small fraction of more iron-enriched stars is
added to the stellar population.

By contrast, the halo of NGC 4889 at radii larger than $~1.2 R_e=18$
kpc is characterized by near-solar metallicities with shallow
gradient, a steep [$\alpha$/Fe] gradient reaching solar values at 60
kpc radius, and old ages (9-13 Gyr, possibly decreasing towards the
outermost radii). The old ages and lower but still enhanced
[$\alpha$/Fe] indicate that the stars in the halo were formed at early
times, but over longer time scales than the stars in the core, out of
gas that was already iron-enriched by exploding type Ia supernovae
\citep{Matteucci94,Bekki+99}.

While galactic winds powered by supernova and/or AGN feedback from the
collapsing central galaxy could explain early truncation of the star
formation in the halo due to removal of the gas, they cannot explain
the longer star formation timescale for the halo regions with respect
to the inner galaxy. On the other hand, the properties of the stellar
population in the halo of NGC4889 are consistent with the idea that
these stars were accreted from shredded satellite galaxies
\citep{Abadi+06,Naab+09}.

These satellites could have formed in lower density regions and
sustained their star formation histories for longer times, until
infall into denser environments and interactions truncated star
formation.  The more extended star formation would enable SN Ia to
contribute iron to the gas and allow the build-up of near-solar
metallicities and $\alpha$-enhancement.  The merger of these
satellites with NGC 4889 could have happened long after the star
formation was truncated.  The total luminosity in the NGC~4889 halo is
$L_{HALO}\sim 10^{11} L_{B,\odot}$, integrating the surface photometry
for $R=1.2-4 R_e$.  This light is equivalent to a few Milky-Way size
galaxies. The merging of a few such satellites with a BCG is
expected in merger trees computed for BCGs in massive clusters
like Coma \cite[e.g.][]{Murante+07}.

The bi-modal nature of the stellar populations in the core ($R<18$
kpc) and halo (18 kpc $< R < 60$ kpc) of NGC 4889 is also consistent
with recent results on the size evolution of early-type galaxies with
redshift \citep[e.g][]{Daddi+05,Trujillo+07,Cimatti+08}, such that
ETGs at $z\sim1$ ($z\sim2$) have sizes a factor of 2 (3-5) smaller
than ETGs with similar mass today
\citep{vanderWel+08,vanDokkum+08,Saracco+09}. Recently,
\citet{vanDokkum+10} find from stacked rest-frame R-band images that
massive ETGs have nearly constant mass within 5 kpc with redshift, but
increase their envelope mass by a factor $\sim 4$ since $z=2$, with
effective radius evolving as $R_e\propto (1+z)^{-1.3}$. Scaling the
present $R_e$ of NGC 4889 with this relation would predict $R_e\sim
6.2$ kpc at $z=1$.
An estimate for the size of the inner [$\alpha$/Fe]-enhanced
  population in this galaxy is the radius enclosing half of the
  current luminosity within 18 kpc, obtained by truncating the surface
  brightness profile \citep{Thomas+07} at that radius; this is also
  $\sim$ 6 kpc.
The consistency of these numbers suggests that we may have
found local stellar population signatures of the observed ETG size
evolution.

In general, BCG galaxies have a range of [Z/H] gradients and
[$\alpha$/Fe] (see Introduction), thought to arise from
the relative influence of the early collapse component and subsequent
mergers \citep{Kobayashi04,DeLucia+07}. In NGC 4889, a relatively
large ($\sim$50\%) fraction of the galaxy light appears to have been
involved in the early collapse. Because of the steep [Z/H] gradient and high
[$\alpha$/Fe] this galaxy may be a particularly good case for
distinguishing the different core and halo populations. It will be
important to obtain similarly extended [$\alpha$/Fe] and [Z/H]
profiles for a larger sample of bright ETGs and to connect their
analysis with the properties of the high-redshift ETG population.

\section*{Acknowledgments}
The authors thank T.~Hattori, S.~Okamura, G.~Ventimiglia, N.~Yasuda
for support and help with the observations and preparations,
R. Bender, K. Freeman, H. Kuntschner, C. Maraston, F. Matteucci,
L. Morelli, R. Saglia, D. Thomas and J. Thomas for useful discussions
in the course of this project, and the anonymous referee for helpful
comments.

\begin{table*}
\centering
\caption{Line strength indices and stellar population parameters for NGC 4889, averaged
in radial bins.}
\begin{tabular}{l c c c c c c c c c }
\noalign{\smallskip}
\hline
\noalign{\smallskip}
$R$  &    \hb         &    \mgb       &      [MgFe]  &      $<$Fe$>$  &      Age               &    [Z/H]            & [$\alpha$/Fe]       &   GAL/BKG             &   $\Delta R$   \\ 
(kpc)&  (\AA)         &    (\AA)      &     (\AA)     &  (\AA)         &    [Gyr]               &                     &                    &                       &     [arcsec]   \\
(1)  &      (2)       &   (3)         &    (4)        &   (5)          &  (6)                   &          (7)        &        (8)         &              (9)      &       (10)     \\
\noalign{\smallskip}
\hline
\noalign{\smallskip}
0.14 &	1.64$\pm$0.15 & 5.46$\pm$0.17 &	4.19$\pm$0.11 &	3.22$\pm$0.13  &  $5.0_{-2.0}^{4.4}$	& $0.63_{-0.13}^{0.17}$&	$0.30_{-0.06}^{0.06}$ &   --                & --               \\
0.43 &	1.61$\pm$0.16 & 5.35$\pm$0.17 &	4.05$\pm$0.11 &	3.07$\pm$0.13  &  $7.8_{-4.0}^{3.2}$	& $0.50_{-0.10}^{0.13}$&	$0.32_{-0.06}^{0.06}$ &   --                & --              \\
0.84 &	1.57$\pm$0.05 & 5.46$\pm$0.08 &	4.07$\pm$0.06 &	3.03$\pm$0.08  &  $8.6_{-1.6}^{1.6}$	& $0.50_{-0.03}^{0.07}$&	$0.34_{-0.02}^{0.02}$ &   --                & --              \\
1.73 &	1.55$\pm$0.06 & 5.54$\pm$0.08 &	4.02$\pm$0.06 &	2.92$\pm$0.08  &  $9.4_{-2.4}^{1.6}$	& $0.47_{-0.07}^{0.03}$&	$0.38_{-0.04}^{0.02}$ &   --                & --              \\
3.67 &	1.66$\pm$0.12 & 5.36$\pm$0.11 &	3.77$\pm$0.07 &	2.65$\pm$0.08  &  $8.6_{-4.0}^{3.6}$	& $0.40_{-0.07}^{0.13}$&	$0.44_{-0.04}^{0.08}$ &   --                & --              \\
8.74 &	1.67$\pm$0.27 & 4.66$\pm$0.30 &	3.69$\pm$0.15 &	2.91$\pm$0.14  &  $8.2_{-5.6}^{3.2}$	& $0.33_{-0.13}^{0.23}$&	$0.22_{-0.12}^{0.06}$ &   --                & --              \\
     &                &               &     [MgFe]'   &                &                        &                     &                      &                     &                \\ 
\noalign{\smallskip}
7    &	1.32$\pm$0.12 & 4.68$\pm$0.17 &	3.48$\pm$0.12 &	2.57$\pm$0.11  &  $>12.2$       	& $0.10_{-0.07}^{0.06}$&	$0.30_{-0.07}^{0.04}$   &  4.7              &  8$"$  (1) \\
12   &	1.33$\pm$0.13 & 4.48$\pm$0.12 &	3.33$\pm$0.07 &	2.47$\pm$0.08  &  $>12.2$	        & $0.01_{-0.05}^{0.03}$&	$0.33_{-0.04}^{0.04}$   &  1.9              &  7$"$  (6) \\
18   &	1.56$\pm$0.12 & 4.26$\pm$0.14 &	3.07$\pm$0.09 &	2.24$\pm$0.09  &  $>12.2$               &$-0.09_{-0.04}^{0.04}$&	$0.40_{-0.05}^{0.08}$   &  0.9              & 13$"$  (4) \\
32   &	1.71$\pm$0.16 & 3.93$\pm$0.16 &	3.03$\pm$0.10 &	2.48$\pm$0.12  &  $10.80_{-4.95}^{1.65}$	&$-0.07_{-0.06}^{0.10}$& $0.23_{-0.10}^{0.07}$   &  0.4              & 30$"$  (5) \\
59   &	1.77$\pm$0.24 & 3.16$\pm$0.32 &	2.79$\pm$0.17 &	2.52$\pm$0.16  &  $9.70_{-4.95}^{3.30}$	&$-0.17_{-0.09}^{0.12}$&	$0.05_{-0.11}^{0.11}$   &  0.2              & 32$"$  (4) \\
\noalign{\smallskip}
\hline
\noalign{\smallskip}
\end{tabular}
\label{tab:models}
\begin{minipage}{18cm}
  Notes: Upper part of the table based on data from
    \citet{Mehlert+00,Corsini+08}; lower part based on data from Paper
    I. Values for literature data refer to the case in which the Mehlert
  et al.\ \hb\ values are offset-corrected by 0.32\AA, and provide
  [MgFe] rather than [MgFe]' (see Section \ref{sec:observations} for
  details). Col.~1: Mean distance of the radial bins from the galaxy center,
  projected along the major axis.  Cols.~2-5: average values for
  indices and errors in the bins.  Cols.~6-8: Corresponding best fit
  age, metallicity and $\alpha$-enhancement with errors. 
    Col.~9: ratio between counts from galaxy and background (both
    sky and scattered light), collected from all pieces of
    the slit that contribute to that bin. These counts are obtained
    in the wavelength range 5000 \AA\ -- 5150 \AA, which is free from
    intense spectral lines, and therefore representative of the
    galaxy and background continua. Col.~10: total radial extent covered
    by the bin. The number of slit portions (data points
    in Fig. \ref{fig:indices}) used for this bin
    is given in parentheses.
\end{minipage}
\end{table*}

\label{lastpage}

\bibliography{coccato10}

\end{document}